\documentclass{emulateapj}

\begin{document}

\title{Shapes of Stellar Systems and Dark Halos from Simulations of
  Galaxy Major Mergers}
\author{Gregory S. Novak,\altaffilmark{1} 
  Thomas J. Cox,\altaffilmark{2}
  Joel R. Primack,\altaffilmark{3}
  Patrik Jonsson\altaffilmark{3},
  Avishai Dekel\altaffilmark{4}}

\altaffiltext{1}{UCO/Lick Observatories, University of California,
  1156 High Street, Santa Cruz, CA 95064; novak@ucolick.org}
\altaffiltext{2}{Harvard-Smithsonian Center for Astrophysics, 60
  Garden Street, Cambridge, MA 02138; tcox@cfa.harvard.edu}
\altaffiltext{3}{Department of Physics, University of California,
  1156 High Street, Santa Cruz, CA 95064; joel@scipp.ucsc.edu, patrik@ucolick.org}
\altaffiltext{4}{Racah institute of Physics, The Hebrew University,
  Jerusalem 91904, Israel; dekel@phys.huji.ac.il}

\shorttitle{SHAPES OF SIMULATED GALAXY MERGER REMNANTS}
\shortauthors{NOVAK et. al.}

\begin{abstract}  
  Using a sample of 89 snapshots from 58 hydrodynamic
  binary galaxy major merger simulations, we find that stellar
  remnants are mostly oblate while dark matter halos are mostly
  prolate or triaxial.  The stellar minor axis and the halo major axis
  are almost always nearly perpendicular.  This can be understood by
  considering the influence of angular momentum and dissipation during
  the merger. If binary mergers of spiral galaxies are responsible for
  the formation of elliptical galaxies or some subpopulation thereof,
  these galaxies can be expected to be oblate and inhabit their
  halos with the predicted shapes and orientations.  These predictions
  are relevant to observational studies of weak gravitational
  lensing, where one must stack many optically aligned galaxies in
  order to determine the shape of the resulting stacked mass
  distribution.  The simple relationship between the dark and luminous
  matter presented here can be used to guide the stacking of galaxies
  to minimize the information lost.  
\end{abstract}


\maketitle

\section{Introduction}
The shapes and mass profiles of dark matter halos from cosmological
$N$-body simulations have long been studied \citep[][and references
therein]{dubinski:91, navarro:96, allgood:06}.  Cosmological
simulations still lack sufficient resolution to track the shape and
orientation of galaxies within their dark matter halos.  There is no
reason to believe that the shapes of galaxies and dark matter halos
should be similar.  It has only recently become feasible to perform
large suites of high-resolution binary galaxy merger simulations
\citep{naab:03, cox:04, cox:05, robertson:06}, and we here use such
simulations in order to study the shapes of the resulting galaxies and
their host halos statistically.

Observationally, the intrinsic shapes of elliptical galaxies have
remained elusive.  It has long been known that there seem to be at
least two classes of elliptical galaxies: massive, anisotropic
galaxies and lower mass, oblate rotators \citep{bender:88, bender:92}.
However, allowing the possibility of triaxiality leads to 
degeneracies in deprojection \citep{franx:91}.  
\citet{alam:02} and \citet{vincent:05} have used Sloan Digital Sky
Survey (SDSS) data to conclude that not all elliptical galaxies can be
oblate.

The relative orientations of galaxies and their dark halos is relevant
to studies of weak gravitational lensing.  Observers stack many images
of galaxies in order to use the average deformation of the shapes of
background galaxies to infer properties of the foreground mass
distribution.  It is important to stack galaxies coherently in order
to build up a detectable signal.  The model presented here represents
a physically well-motivated \textit{Ansatz} to help interpret the
results of weak lensing observations.  Section \ref{sec:methods} gives
a description of the galaxy merger simulations and our method of
determining the shape of merger remnants, \S \ref{sec:results} gives
our results, and \S \ref{sec:conclusions} summarizes our conclusions.

\section{Methods}
\label{sec:methods}
We analyze the shapes of 89 snapshots from 58 of these simulations.
We study two samples of simulations.  One is the ``G''
series, which consists of major and minor mergers with progenitor
spiral galaxies typical of the nearby universe and spanning a factor
of 40 in baryonic mass and 20 total mass.  In order to reduce the
dependence on the progenitor galaxy model, here we only consider major
mergers with mass ratios of 1:1 (G3-G3, G2-G2, G1-G1, and G0-G0) and
roughly 3:1 (G3-G2, G2-G1, and G1-G0).  We also analyze the ``Sbc''
series of merger simulations, which are 1:1 major mergers of massive,
gas-rich spirals using a variety of different orbits and orientations.

\begin{figure*}[t]
\centering
\plottwo{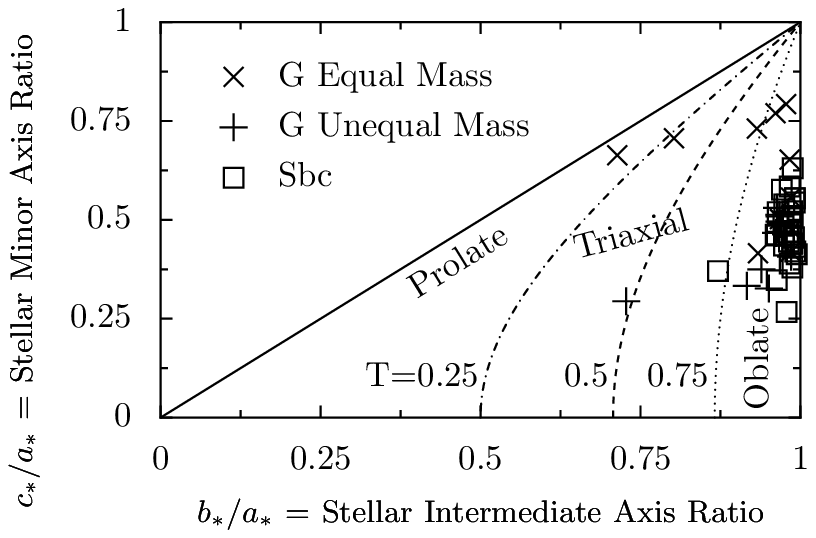}{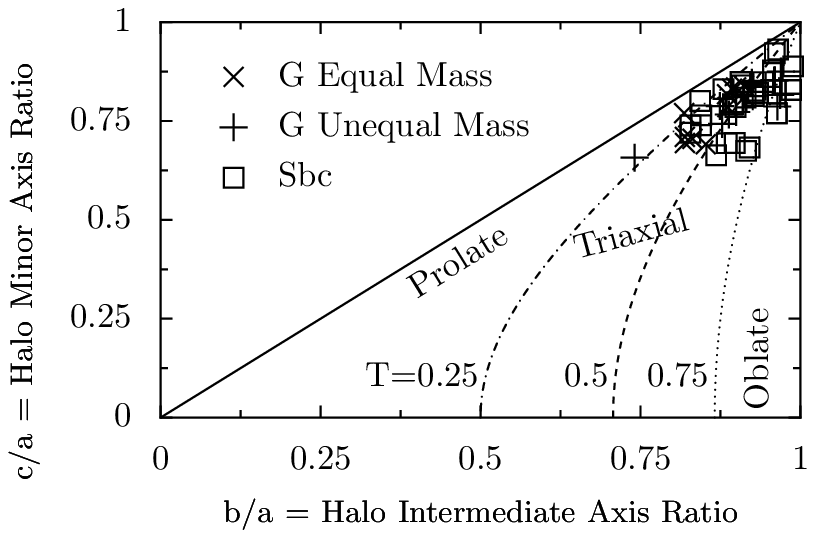}
\caption{
    Shapes of the luminous and dark components of simulated merger
    remnants.  \textit{Left}: the intermediate-to-major axis ratio
    vs. the minor-to-major axis ratio for stars.  \textit{Right}: Same
    as left panel, but for dark matter halos.  Objects near the 
    diagonal line are prolate spheroids, objects near $b/a=1$ are
    oblate spheroids, and objects in between are triaxial.  Dotted
    lines indicate constant triaxiality $T$.  Most
    stellar remnants are oblate with $\epsilon=0.5$.  
    Most dark matter halos are either prolate or
    triaxial.}
  \label{fig:shapes}
\end{figure*}

To calculate the shape of a merger remnant, we iteratively diagonalize
a moment of inertia tensor using an ellipsoidal window \citep{dubinski:91}:
\begin{equation}
  M_{ij} = \Sigma_k m_k r_{i,k} r_{j,k}
  \label{eq:moment-of-inertia}
\end{equation}
where $r_{i,k}$ is the position vector, $i,j$ refer to coordinates, and $k$
refers to particle number.  
The triaxial radius is given by \citet{franx:91}:
\begin{equation}
  \zeta = \sqrt{x^2/a^2 + y^2/b^2 + z^2/c^2}
  \label{eq:triaxial-radius}
\end{equation}
where $a$, $b$, and $c$ are the major, intermediate, and minor axis
lengths, respectively.  The sum over $k$ includes all particles for
which $r$ lies within the ellipsoid $\zeta = 1$.  The iteration is
started with a spherical window ($a=b=c=$ baryonic half-mass radius),
and after each iteration $a$,$b$, and $c$ are scaled so that half of
the baryonic mass is enclosed.  The result does not appreciably change
if equation (\ref{eq:moment-of-inertia}) is modified to include
$\zeta^2$ in the denominator.  Using a spherical window rather than an
ellipsoidal one results in systematically larger axial ratios but does
not change the main result.

Three-dimensional shapes of galaxies can be quantified with the
triaxiality parameter $T=(a^2-b^2)/(a^2-c^2)$.  We call an object
oblate, triaxial, or prolate if $T<0.25$, $0.25<T<0.75$, or $0.75<T$,
respectively.  Shapes of galaxies can also be quantified by
ellipticity $\epsilon=1-b/a$.  Ellipticities are most often used to
describe two-dimensional shapes; we occasionally refer to the
three-dimensional ellipticity of perfectly prolate or oblate ($T=0$ or
1) objects since there is no ambiguity about the use of the
equation.

Simulations were performed using the entropy-conserving version of the
SPH code GADGET \citep{springel:01,springel:02} with a gravitational
smoothing length of 100 pc.  The progenitor galaxies have baryonic
masses from $1.6\times10^{9}$ to $2\times10^{11} M_\odot$, gas
fractions between 20\% and 70\%, consist of $\sim$100,000 particles,
and use a parameterization of star formation feedback from supernovae
tuned to match the empirical Schmidt law \citep{kennicutt:98}.
\citet{cox:04} and \citet{cox:05} contain further information about the
simulations.

\begin{figure}[b]
\centering
\plotone{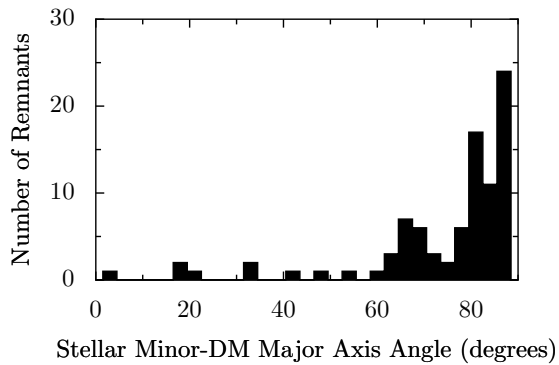}
  \caption{
    The angle between the \textit{minor} axis of the stars and the
    \textit{major} axis of the dark matter.  The stellar material is
    mostly oblate, the dark matter halos are mostly prolate, and the
    ``preferred'' axes of the two shapes are nearly always
    perpendicular.}
  \label{fig:orientation}
\end{figure}

\section{Results}
\label{sec:results}
\begin{figure}[b]
\centering
\plotone{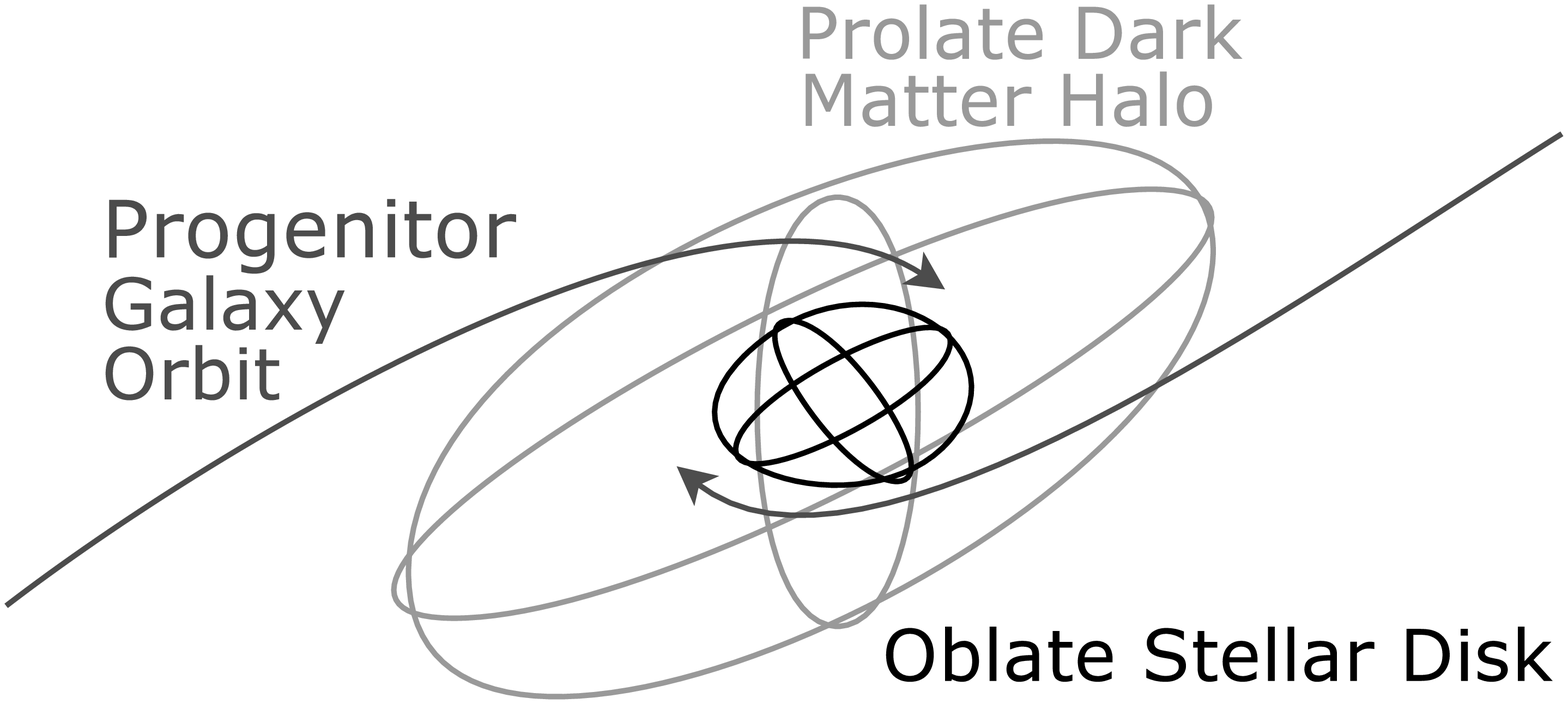}
\caption{
    The physical interpretation of Figs. \ref{fig:shapes} and
    \ref{fig:orientation} in terms of angular momentum and
    dissipation.  The total angular momentum in a merger simulation is
    usually dominated by the orbital angular momentum of the two
    galaxies.  As the galaxies merge, both the luminous and dark
    components acquire angular momentum from the orbits of the
    progenitors.  Their velocity dispersion increases along the axis
    parallel to the direction of approach, leading to an anisotropic
    velocity dispersion tensor.  Gas in the simulation cools while
    largely conserving angular momentum so that it spins up to the
    point where the shape of the resulting stellar system is
    determined by rotation, not velocity dispersion anisotropy.
    Meanwhile, the dark matter cannot cool, so its shape is determined
    by velocity dispersion anisotropy.  Therefore, the stellar system
    is oblate, dark matter halos are prolate, and the angle between
    the ``preferred'' axes of these two shapes is $\simeq$90$^\circ$.}
  \label{fig:diagram}
\end{figure}

Figure \ref{fig:shapes} illustrates that most stellar remnants are
oblate, while the dark matter halos in which they reside are mostly
prolate or triaxial.  Figure \ref{fig:orientation} shows
that the short axis of the stellar system and the long axis of the
dark matter halo are almost always nearly perpendicular.  This can be
understood simply in terms of angular momentum and dissipation, as
shown in Figure \ref{fig:diagram}.

This model helps interpret the findings from studies of
weak gravitational lensing.  \citet{hoekstra:04} find that the
ellipticity of dark halos is
$0.77^{+0.18}_{-0.21}$ times the ellipticity of the light (i.e., halos
are somewhat less flattened than galaxies), assuming that the two are
aligned.  According to our result, elliptical galaxies would either
show an elliptical halo (if the long axis of the prolate halo is in
the plane of the sky) or a circular halo (if the long axis of the halo
is pointed toward the observer).  Thus, the flattening of the dark
matter would follow that of the luminous matter, in agreement with
these observations.  

\begin{figure}[!t]
\centering
\plotone{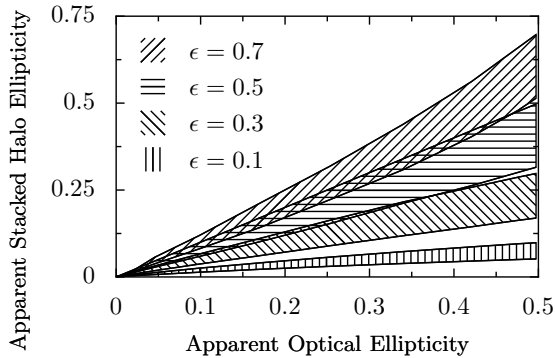}
\caption{    
    Consequences of our model for
    weak gravitational lensing measurements where one must stack many
    optically aligned galaxies in order to detect flattening of the
    dark matter halo.  The plot shows the apparent ellipticity of the
    halo mass surface density vs. the apparent optical ellipticity
    of the stacked galaxies.  Each shaded region shows the range of
    possible apparent ellipticities given an \textit{intrinsic,
      three-dimensional} halo ellipticity.  The lower bound of each
    shaded region is given by assuming a prolate halo and the upper
    bound by assuming an oblate halo.  This is the result one would
    obtain if all galaxies followed the trend noted in this Letter and
    one stacked galaxies only with a \textit{given} optical ellipticity.
    The observed halo ellipticity goes to zero for apparent
    optical ellipticities near zero because the galaxies cannot be
    oriented so that the stacking is coherent, even though each
    individual halo will have a nonzero projected ellipticity.
    The observed halo ellipticity only
    equals the three-dimensional ellipticity when halos are
    intrinsically oblate and the galaxies are viewed edge
    on; otherwise the flattening is underestimated.}
  \label{fig:lensing-1}
\end{figure}

\begin{figure*}[!htb]
\centering
\plottwo{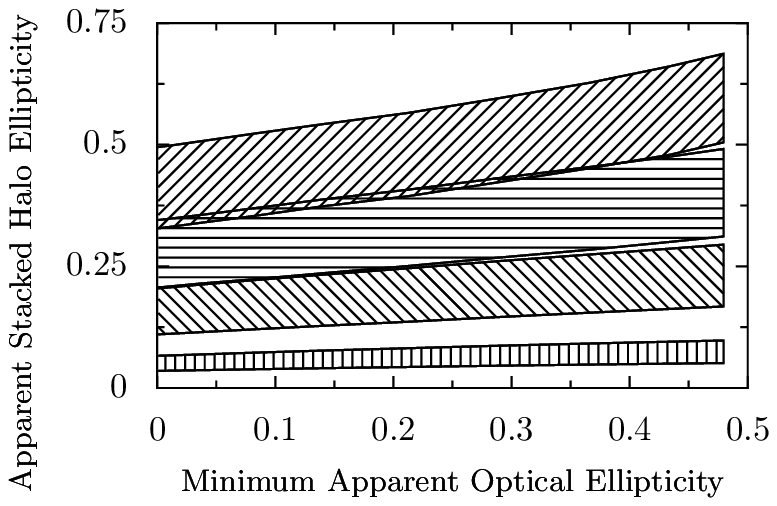}{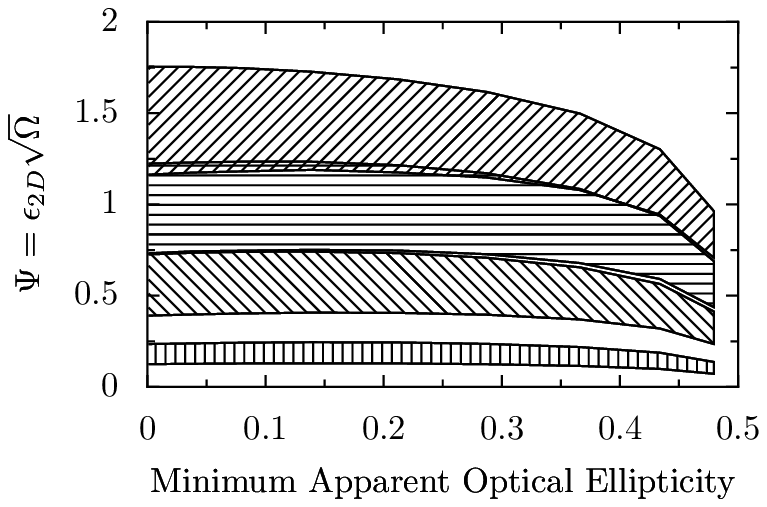}
\caption{Weak gravitational lensing measurements when all galaxies
    with optical ellipticities greater than some value are included in
    the stacking.  The shaded regions are defined as in
    Fig. \ref{fig:lensing-1}.  In both panels the $x$-axis is the
    minimum apparent optical ellipticity of galaxies included in the
    stacking.  \textit{Left}: The $y$-axis shows the apparent
    ellipticity of the stacked halo mass surface density.  This is the halo
    ellipticity that would be measured if one stacked \textit{all}
    galaxies with optical ellipticity greater than the given value.
    \textit{Right}: Effect of different optical ellipticity
    requirements on the S/N ratio of the halo ellipticity
    measurement.  Increasing the minimum optical ellipticity increases
    the signal (as seen in the right panel) but also increases the
    noise by reducing the number of galaxies included in the stack.
    The $y$-axis is $\Psi$ defined by eq. (\ref{eq:psi}).  
    For small optical ellipticity cuts,
    the increase in signal is almost exactly canceled by the decrease
    in available solid angle, meaning that nothing is lost (or gained)
    by removing round galaxies from the stack.  In reality, there is
    an advantage to removing low-ellipticity galaxies since their
    position angle is ill-defined.  (This analysis assumes no error in
    position angle.)  Above a minimum ellipticity of $\sim0.25$ the
    quality of the measurement declines rapidly because of smaller
    solid angle of available viewing directions.}
  \label{fig:lensing-2}
\end{figure*}

The interpretation of the \citeauthor{hoekstra:04}
data is complicated by the inclusion of spiral as well as elliptical
galaxies in the sample.  \citet{mandelbaum:06a, mandelbaum:06b} have
done a similar study using SDSS galaxies separated by Hubble type and
found that the projected halo shapes for elliptical galaxies are
aligned with the projected stellar shapes, in agreement with
\citeauthor{hoekstra:04}  Finally, the projected positions of
satellite galaxies also seem to indicate that the projected shapes of
elliptical galaxies and halos are aligned \citep{sales:04,
  brainerd:05, yang:06}.

Weak lensing studies necessarily underestimate the flattening of dark
matter halos. Figures \ref{fig:lensing-1} and \ref{fig:lensing-2}
quantify this by simulating the weak gravitational lensing
observations.  Given assumptions about the three-dimensional shapes
and mass profiles of galaxies and their halos and a scheme for
combining many galaxies into a single mass surface density, these two
figures show shapes of the projected halo mass surface densities.
They allow observers to translate their two-dimensional
measurements to a range of possibilities for the three-dimensional
structure of dark matter halos.

The hydrodynamic simulations discussed here do not represent a
cosmologically unbiased sample, so they are \textit{not} used as input
to the simulated lensing observations.  Instead we adopt a slightly
idealized version of the correlation between halos and galaxies noted
in this Letter.  Nearly all of the baryonic components of the
simulated galaxies are close to 2:1 oblate spheroids, so we assume
that all early-type galaxies are so described.  Thus, there is a
simple mapping between viewing angle and optical ellipticity.  We
assume all galaxies follow the correlation between halos and galaxies
noted here and that the halo mass density is given by a triaxial
Navarro-Frenk-White profile: $\rho=\rho_0/(\zeta/r_s)(1+\zeta/r_s)^2$,
where $\rho$ is the mass density and $\rho_0$ is a constant
\citep{navarro:96, jing:02}.

\citet{contopoulos:56} showed that for a triaxial ellipsoid with
constant three-dimensional axis ratios, the contours of constant
projected surface density are ellipses with constant ellipticity and
position angle, independent of the radial density profile.  We only
use the ellipticity and position angle of the baryonic component, so
the radial profile of the baryons does not matter.  The
\citet{contopoulos:56} analysis does not apply to the stacked dark
matter halos, so Figures \ref{fig:lensing-1} and \ref{fig:lensing-2}
depend on the radial density distribution of the halos.  In practice
the difference is not large.

To simulate weak lensing measurements, we align the projected mass
distributions based on projected light distributions, stack the
projected halo mass distributions, and fit an ellipse to the halo mass
surface density distribution where the area of the ellipse is
constrained to equal $\pi (3 r_s)^2$.  This size
for the ellipse is motivated by the approximate radius at which weak
lensing observations are sensitive to the halo shape (M. J. Hudson 2006,
personal communication).  The stacking either assumes a given
inclination of the optical galaxy, averaging over the azimuthal angle
(as in Fig. \ref{fig:lensing-1}), or assumes that some \textit{minimum}
optical ellipticity is required to be included in the stack, averaging
over the portion of the unit sphere that gives rise to sufficient
optical ellipticities (as in Fig. \ref{fig:lensing-2}).

Figure \ref{fig:lensing-1} shows that galaxies with low optical
ellipticities will have low halo ellipticities because there is no
preferred axis to use to stack galaxies.  The only situation where the
projected halo ellipticity equals the three-dimensional halo
ellipticity is when all stacked galaxies are viewed edge-on and halos
are intrinsically oblate.  Flattening is underestimated in all other
cases.  

Figure \ref{fig:lensing-2} shows the result of the more realistic
scenario where all galaxies with optical ellipticities greater than
some value are included in the stack.  This allows one to transform
projected ellipticities to three-dimensional ellipticities.  For
example, if an observer sets the minimum optical ellipticity to 0.2
and measures a stacked halo ellipticity of 0.25, one can conclude that
the three-dimensional ellipticity of halos is either 0.3 (for oblate
halos), 0.5 (for prolate halos), or somewhere in between.

As one enforces tighter constraints on the optical ellipticity, the
halo ellipticity goes up, but the cost is that fewer galaxies will
make it into the stack.  Under simple assumptions, one can estimate
the signal-to-noise ratio (S/N) of the halo ellipticity measurement to be
\begin{equation}
  (\mbox{S/N})_{\mbox{tot}} = 
  \epsilon_{\mbox{2D}} \sqrt{ \Omega N_{\mbox{tot}}} / \sigma_1
  \label{eq:s/n}
\end{equation}
where $\epsilon_{\mbox{2D}}$ is the apparent ellipticity of the stacked halo
mass surface density, $\Omega$ is the solid angle of viewing angles
for which a galaxy will be included in the stack, $N_{\mbox{tot}}$ is the
total number of galaxies in the survey, and $\sigma_1$ is the error on
the halo ellipticity when only \textit{one} galaxy is used.  We
define $\Psi$ as the part of this expression which depends on the
signal and the available solid angle:
\begin{equation}
  \Psi = \epsilon_{\mbox{2D}} \sqrt \Omega 
  \label{eq:psi}
\end{equation}
Figure \ref{fig:lensing-2} thus also allows observers to estimate the
quality of their measurement given the size of their survey and an
estimate of the one-galaxy error on the halo ellipticity.  In reality,
observers do not know $\Omega$, but they could estimate it from the
from the minimum ellipticity of galaxies in their sample as long as
our assumption that the stellar remnants are 2:1 oblate spheroids is not
far wrong.

\section{Discussion}
\label{sec:conclusions}
We have analyzed the three-dimensional shapes of galaxies and dark
matter halos resulting from more than 100 simulations of gas-rich
galaxy mergers.  Stellar remnants are nearly all oblate, with a few
examples of triaxiality in the most gas-poor mergers.  Dark matter
halos are either prolate or triaxial, and the short axis of the baryons
is perpendicular to the long axis of the dark matter.  All of these
facts can be understood in terms of the effects of angular momentum
and dissipation during the merger.  If there is a class of elliptical
galaxies that were formed by gas-rich binary galaxy mergers, they can
be expected to display these characteristics.  

Real galaxies in a $\Lambda$CDM universe are thought to have experienced
many mergers over the course of their history, and these multiple
mergers can be expected to weaken the relationship between the shapes
of galaxies and their halos described here.  The extent to which the
effects of large-scale structure, such as mass accretion along
filaments, tend to preserve the relationship between galaxies and
their halos is an interesting and open question.

\acknowledgments

G. S. N. was supported by the Krell Institute through the
Computational Science Graduate Fellowship Program.  Computing
resources were provided by the UCSC Beowulf cluster UpsAnd and NERSC.
We thank Andreas Burkert, S. M. Faber, Mike Hudson, and Laura Parker
for useful discussions.

\bibliographystyle{apj}

\bibliography{ms}

\end{document}